\documentclass[10pt,journal]{IEEEtran}
\usepackage{amsmath,amsfonts}
\usepackage{algorithmic}
\usepackage{algorithm}
\usepackage{array}
\usepackage{booktabs}
\usepackage[caption=false,font=normalsize,labelfont=sf,textfont=sf]{subfig}
\usepackage{textcomp}
\usepackage{stfloats}
\usepackage{url}
\usepackage{verbatim}
\usepackage{graphicx}
\usepackage{cite}
\usepackage{graphicx}
\usepackage{subfloat}  
\usepackage{cleveref}
\usepackage{multirow}
\usepackage{diagbox}

\begin{document}
	\begin{sloppypar}
	\title{Deep Quality Assessment of Compressed Videos: A Subjective and Objective Study}
	
	\author{Liqun~Lin,~\IEEEmembership{Member,~IEEE,}
			Zheng~Wang, 
			Jiachen~He, 
			Weiling Chen,~\IEEEmembership{Member,~IEEE,}
			
			Yiwen~Xu,~\IEEEmembership{Member,~IEEE,}
		    and~Tiesong Zhao,~\IEEEmembership{Senior Member,~IEEE}

\thanks{This work was mainly supported by the Natural Science Foundation of China under Grant 62171134 and Grant 61901119.  It was also supported by the Fujian Provincial Education Department under Grant JAT200024. \it{(\it Corresponding~author: Yiwen~Xu.)}}
\thanks{The authors are with the Fujian Key Lab for Intelligent Processing and Wireless Transmission of Media Information, College of Physics and Information Engineering, Fuzhou University, Fuzhou, Fujian 350116, China (e-mails: \{lin\_liqun, n191120078, 061900119, weiling.chen, xu\_yiwen, t.zhao\}@fzu.edu.cn).
		}

}
	\markboth{Journal of \LaTeX\ Class Files,~Vol.~14, No.~8, August~2022}%
	{Shell \MakeLowercase{\textit{et al.}}: A Sample Article Using IEEEtran.cls for IEEE Journals}

	\maketitle

\begin{abstract}
 In the video coding process, the perceived quality of a compressed video is evaluated by full-reference quality evaluation metrics. However, it is difficult to obtain reference videos with perfect quality. To solve this problem, it is critical to design no-reference compressed video quality assessment algorithms, which assists in measuring the quality of experience on the server side and resource allocation on the network side. Convolutional Neural Network (CNN) has shown its advantage in Video Quality Assessment (VQA) with promising successes in recent years. A large-scale quality database is very important for learning accurate and powerful compressed video quality metrics. In this work, a semi-automatic labeling method is adopted to build a large-scale compressed video quality database, which allows us to label a large number of compressed videos with manageable human workload. The resulting Compressed Video quality database with Semi-Automatic Ratings (CVSAR), so far the largest of compressed video quality database. We train a no-reference compressed video quality assessment model with a 3D CNN for SpatioTemporal Feature Extraction and Evaluation (STFEE). Experimental results demonstrate that the proposed method outperforms state-of-the-art metrics and achieves promising generalization performance in cross-database tests. The CVSAR database and STFEE model will be made publicly available to facilitate reproducible research.
\end{abstract}
\section{Introduction}
\IEEEPARstart{T}{he} perceptual quality of a compressed video is an extremely important indicator in video coding. With the development of the network, the video traffic shows an explosive growth trend \cite{1}. However, the network bandwidth is limited. How to transmit video better and faster under the limited network resources is a very meaningful topic. The greater the degree of video compression, the faster the transmission of video resources. Popular video compression methods are lossy compression, which directly reduces the perceptual quality of videos \cite{2}. Therefore, a reliable Video Quality Assessment (VQA) method for compressed videos is critical as both a performance indicator and a guidance for further improvement. Peak Signal-to-Noise Ratio (PSNR)\cite{3} and Structural SIMilarity (SSIM) index \cite{4} are the mainstream quality assessment methods. However, the characteristics of human visual characteristics and temporal characteristics between video frames are not considered, the perceptual quality of videos cannot be accurately expressed.

Existing VQA methods can be classified into subjective VQA methods and objective VQA models. Subjective VQA methods are subjective tests by human observers, which is the most reliable evaluation method because the user is the ultimate viewer of videos. However, this method is time-consuming and impractical. Instead, it is often used in the construction of various quality evaluation databases to provide a reliable reference for objective VQA methods. Objective VQA models are guided by subjective scores and predict the quality of videos by automatic algorithms. Objective VQA models are widely used because of their convenience and low cost. 

Numerous objective VQA models have been developed. According to the availability of original reference videos, objective VQA models can be divided into three categories: Full-Reference VQA (FR-VQA), Reduce-Reference VQA (RR-VQA) and No-Reference VQA (NR-VQA). For FR-VQA, PSNR and SSIM are the most common full-reference evaluation algorithms. In addition, Fu {\it {et al}}. \cite{5} utilized CNN to obtain the perceptual features of each frame. Then, combined with a self-attention module, frame-level features were fused to video-level features, achieving video quality prediction. Li {\it {et al}}. \cite{6} predicted compressed video quality by measuring the frame difference between reference and distorted video frames to measure the relative standard deviation. However, in practical applications, it is difficult to obtain reference videos with perfect quality, so the practical value of such evaluation algorithms is very limited. RR-VQA needs some features of reference videos to participate in predicting video quality, and its practical value is also limited. Reference \cite{7} utilized Gaussian scale mixture model to analyze the conditional entropy of reference image and distorted image, and constructed RR-VQA method on spatial scale. Although, the advantage of NR-VQA without any reference information is of great application value \cite{8,9,10}, most of the existing NR-VQA methods have the limitation of artificially extracting features, and are difficult to obtain ideal generalization performance. 

For the perceived quality prediction of compressed videos, video codec technology usually takes NR-VQA as the main evaluation methods, but it is difficult to obtain reference videos with perfect quality. To solve this problem, it is critical to design NR-VQA algorithms for compressed videos. Video encoding often introduce mixed complex artifacts. In addition to common spatial artifacts, temporal artifacts are also degrade user's perceptual experience. The key challenge in VQA is how to effectively learn spatiotemporal features of compressed videos and then map the features to video quality prediction. Convolutional Neural Network (CNN) has shown its advantage in VQA with promising successes in recent years \cite{11,12,13,14}. CNN can jointly learn features and make quality predictors in an end-to-end manner. Despite its advantages, CNN has not been well developed in compressed video quality evaluation, which is limited by the lack of large-scale compressed video quality evaluation databases. Therefore, there is still room for improvement in the performance of data-driven compressed video quality evaluation algorithms.

The performance of deep learning-based VQA models extremely depends on the quality and quantity of training datasets. Although there are several video quality databases of compressed videos have been built, such as LIVE Video Quality Database (LIVE) \cite{15}, LIVE Mobile VQA Database \cite{16}, CSIQ \cite{17}, IVP \cite{18} and WaterlooIVC4K Video Quality Database (WaterlooIVC4K) \cite{19}, {\it {etc}}. However, they are not enough for deep network training in their sample sizes. Among them, WaterlooIVC4K is the largest database, which contains only 1200 videos of 20 different video contents. Therefore, it is imperative to establish a larger compressed video quality database. To this end, the biggest challenge lies in the huge workload of mankind. In this work, we observe an exponential attenuation relationship between the coding parameters of compressed videos and their subjective quality score. It helps us to develop a large database that reduces the workload of manual annotation. Experiments on randomly selected samples prove the high accuracy of the proposed database. Based on this database, we construct a perceptual data-driven NR-VQA model to predict compressed video quality, which is highly related to subjective score. The major contributions of our work are summarized as follows:

1) A large-scale quality database of compressed videos is developed with a novel semi-automatic subjective labeling method, which greatly reduces the workload of manual labeling.

2) A no-reference compressed video quality assessment with a 3D CNN for SpatioTemporal Feature Extraction and Evaluation (STFEE) is proposed.  It results in an end-to-end model that jointly learns perceptually spatiotemporal features of compressed videos and a quality predictor.

3) Superior performance of our method is achieved against state-of-the-art quality prediction. In addition, our algorithm can achieve reasonable performance in cross-database verification, which shows that our algorithm has good generalization and robustness.

The rest of this paper is organized as follows. In Section \uppercase\expandafter{\romannumeral2}, the related work is introduced. In Section \uppercase\expandafter{\romannumeral3}, we elaborate the details of the method and procedure to build the proposed large-scale database. In Section \uppercase\expandafter{\romannumeral4}, we describes the proposed deep network including network structure, feature extraction method and model training. In Section \uppercase\expandafter{\romannumeral5}, the relevant experimental results are given. Finally, this paper is concluded in Section \uppercase\expandafter{\romannumeral6}.
	
\section{RELATED WORK}
Video compression greatly reduces video stream bit rate and improves transmission efficiency. 
Despite its high coding efficiency, the encoded videos often show visually annoying artifacts, which significantly degrade the visual quality experience of end users. In order to evaluate the perceptual quality of compressed videos, numerous perceptual VQA metrics have been proposed to predict the visual quality of videos with FR-VQA \cite{3,4,5,6}, RR-VQA \cite{7} and NR-VQA \cite{11,12,13,14}. Among them, the application field of FR-VQA metric is limited owing to its requirement of unimpaired video source. Thus, RR-VQA and NR-VQA algorithms are preferred.

In addition, database is a crucial element in the field of compressed video quality evaluation. The performance of VQA algorithms has a significant dependence on the quality of databases. Literature \cite{15,19} contain different levels of distortion and different compression methods to form different amounts of video sets. Based on the current popular compressed video quality databases, the existing compressed video quality evaluation algorithms can be divided into three categories: single-distortion quality evaluation algorithms, multi-distortion quality evaluation algorithms and deep learning quality evaluation algorithms.

Single distortion quality evaluation methods are aimed at a specific type of distortion, which have low generalization performance and poor practicability. Choi {\it {et al}}. \cite{20} studied the relationship between temporal visual masking and subjective perception quality for local flicker in compressed videos. Korhonen {\it {et al}}. \cite{21} observed the effect of packet loss and subjective perception in decoded videos. Multi-distortion quality evaluation methods are based on Natural Scene Statistic (NSS) \cite{22}. Mittal {\it {et al}}. \cite{23} standardized the frame difference and fitted the product of the four directions calculated by the standardized coefficient with asymmetric generalized Gaussian distribution. Finally, shape parameters were utilized to regress the perceived quality fraction of videos. Reddy {\it {et al}}. \cite{24} calculated 3D Mean Subtractedand Contrast Normalized coefficients of videos (3D-MSCN). A 3D Gabor filter was designed to fit the asymmetric generalized Gaussian distribution after 3D-MSCN filtering, and the mapping model of shape parameters and perceived quality of visual frequency were trained by Support Vector Regression. In addition, learning-based methods automatically learn the mapping between video features and the perceptual quality, which have been growing steadily \cite{25,26}. Liu {\it {et al}}. \cite{1} established a large number of compressed videos to train deep learning network model. However, the Mean Opinion Score (MOS) of the database was obtained by full reference SSIMplus \cite{11}, which is difficult to guarantee the high correlation between the MOS and objective perception. In order to avoid the problem of lack of data, some scholars utilized transfer learning to evaluate video quality. Li \cite{12} {\it {et al}}. extracted features from the pre-trained image classification neural network ResNet50 to obtain perception characteristics. Chen {\it {et al}}. \cite{13} adopted VGG-16 network to learn the frame-level features of videos, and then obtained the Gaussian distributed features through adversarial learning.
%
		
		

The above great efforts focus on manually extracting features to predict video quality, which have insufficient generalization performance. In addition, although deep learning can solve this problem, due to the lack of large-scale compressed video quality evaluation databases, the development of data-driven compressed video quality evaluation methods is still not perfect. The content perception features extracted by several methods of transfer learning \cite{12,13} are all in the spatial domain, which are obviously not enough for videos. To solve the above problems, it is necessary to construct a large compressed video quality evaluation database and propose a data-driven method to learn their spatiotemporal features. In this paper, a large compressed video quality database is established and a semi-automatic marking method is used to obtain the perceptual quality score of compressed videos quickly. On this basis, the STFEE algorithm is proposed.
\section{Proposed large-scale CVSAR}
The quality evaluation of compressed videos should accurately reflect the subjective perception of human eyes. Therefore, a compressed video quality database with subjective scores is an indispensable factor in the construction of a compressed video quality evaluation algorithm. In general, compressed video databases contain different content videos with different quality levels. In addition, each video with a reliable subjective quality score is the most critical to develop VQA algorithms. At present, there are some quality evaluation databases for compressed videos, such as LIVE Video Quality database, LIVE Mobile VQA Database, CSIQ, IVP and WaterlooIVC4K Video Quality Database. The related information is shown in Table \ref{t1}. 
\begin{table*}[ht]
	\centering\small{
		\caption{Existing Compressed Video Databases\label{t1}}
		\newcommand{\tabincell}[2]{\begin{tabular}{@{}#1@{}}#2\end{tabular}}
		\begin{tabular}{ccccc}
			\specialrule{1.5pt}{3pt}{3pt}
			\tabincell{c}{Database} &\tabincell{c}{Number of\\original videos} & \tabincell{c}{Type of\\compression} &\tabincell{c}{Number of\\compressed videos}&\tabincell{c}{Year of\\presentation}\\
			\specialrule{0.05em}{2pt}{2pt}
			\multicolumn{1}{c}{LIVE Video Quality Database} &\multicolumn{1}{c}{10} & \multicolumn{1}{c}{2} &\multicolumn{1}{c}{80}&\multicolumn{1}{c}{2010}\\
			\multicolumn{1}{c}{LIVE Mobile VQA Database} &\multicolumn{1}{c}{10} & \multicolumn{1}{c}{1} &\multicolumn{1}{c}{40}&\multicolumn{1}{c}{2012}\\
			\multicolumn{1}{c}{CSIQ} &\multicolumn{1}{c}{12} & \multicolumn{1}{c}{3} &\multicolumn{1}{c}{108}&\multicolumn{1}{c}{2012}\\
			\multicolumn{1}{c}{IVP} &\multicolumn{1}{c}{10} & \multicolumn{1}{c}{3} &\multicolumn{1}{c}{50}&\multicolumn{1}{c}{2012}\\
			\multicolumn{1}{c}{WaterlooIVC4K} &\multicolumn{1}{c}{20} & \multicolumn{1}{c}{5} &\multicolumn{1}{c}{1200}&\multicolumn{1}{c}{2019}\\
			\specialrule{1.5pt}{0pt}{0pt}
	\end{tabular}}\\		
\end{table*}

None of the above compressed video databases are sufficient for training deep neural networks. Databases such as LIVE and CSIQ contain a limited number of videos to adequately to train complex networks. Although WaterlooIVC4K contains 1200 videos, it contains only 20 limited video contents. The contents of the videos are not rich enough, and the model is prone to overfitting during the training process.  Large-scale databases are desired to develop learning-based VQA methods for training. In building such a database, the main challenge is how to rate quality scores for a large number of videos. The desired subjective tests are time-consuming and costly. Moreover, the consistency and reliability of subjective ratings are affected by fatigue effects labeling large-scale datasets. Zhao {\it {et al}}. \cite{43} constructed a super-resolution image quality database in a semi-automatic labeling manner, which greatly reduced the cost of database construction. On the labeling method, Zhao {\it {et al}}. observed a functional relationship between the number of samples and image quality. Based on this functional relationship, the current largest super-resolution image quality database is quickly constructed. Inspired by this, we explore the quality variation law of compressed videos and construct a database based on this law to greatly reduce the workload of subjective testing.

\subsection{The Quality Variation Law in compressed videos}
To train learning-based VQA methods, large-scale databases are desired, which will inevitably lead to the explosive growth of subjective testing workload. To solve the problem, a semi-automatic rating mechanism is proposed by observing the mapping relationship between compressed video quality and related compression parameters. Semi-automatic labeling based on the mapping relationship modeling can ensure the reliability of labels and reduce the construction cost of large-scale databases.

\begin{figure}[htpb] 
	\captionsetup{font=small,labelfont=bf}
    \includegraphics[width=0.5\textwidth]{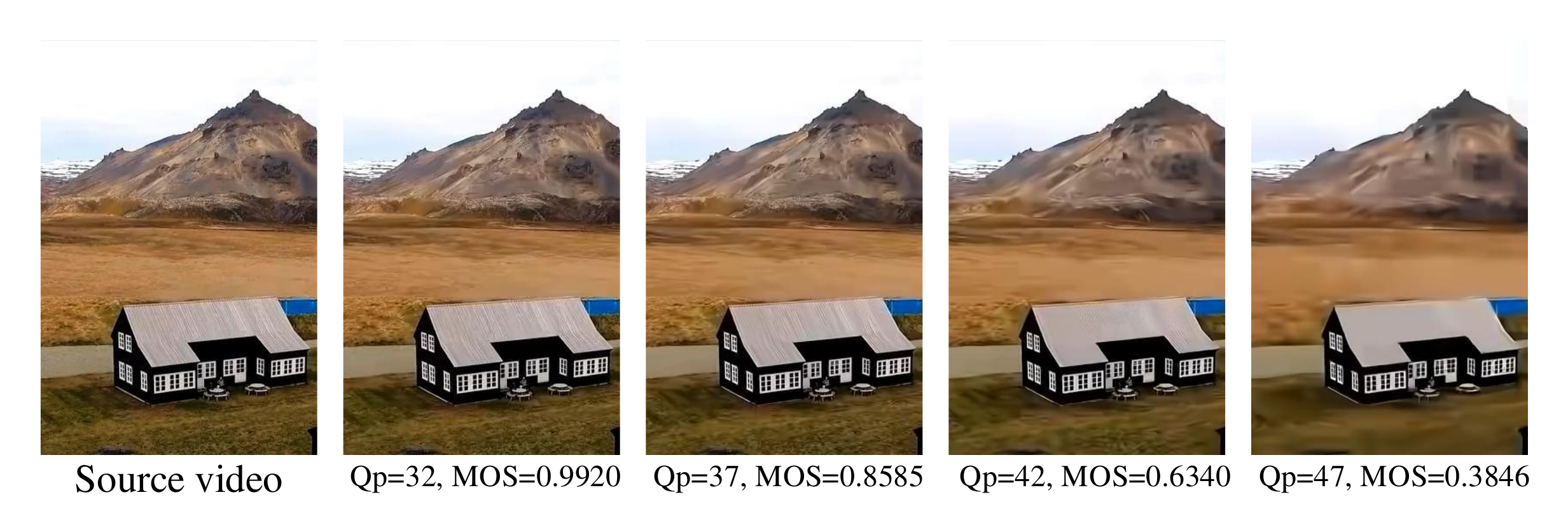}
	\caption{Videos generated by the same compression method with different Qp values.}
	\label{f2} 
\end{figure}

The most important parameter of video compression is the Quantization parameter (Qp). Video perceived quality is affected by its Qp. Fig.\ref{f2} shows the same video compressed by the same compression method with different Qp values. With the increase of Qp value, the loss of details becomes more serious, for example, the roof and background become smoother, thus resulting in lower perceived quality.

\begin{figure*}[tbp]
	\centering
	\subfloat[][] 
	{
		\begin{minipage}{9cm}
			\centering          
			\includegraphics[scale=0.8]{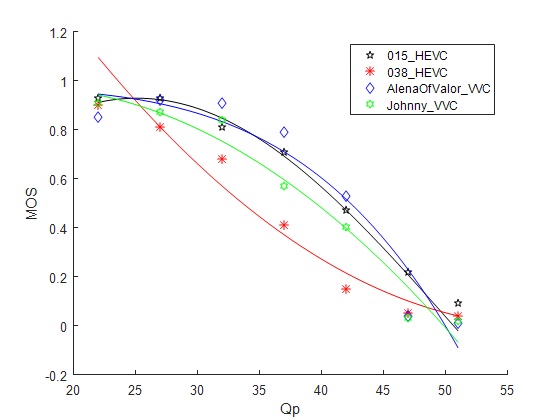}   
			\label{f31}
		\end{minipage}
		
	}
	\subfloat[][] 
	{
		\begin{minipage}{9cm}
			\centering      
			\includegraphics[scale=0.8]{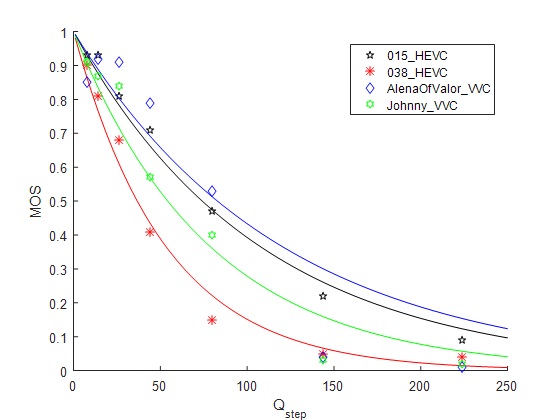}
			\label{f32}
		\end{minipage}
		
	}
	\captionsetup{font=small,labelfont=bf}
	\caption{Relationship between MOS,Qp and $Q_{\rm step}$. (a) Relationship between MOS and Qp, (b) Relationship between MOS and $Q_{\rm step}$.} 
	\label{f3}
\end{figure*}
To verify the relationship between Qp value and video quality, we select different scene videos encoded with different Qp values, and then conduct subjective experiments to obtain the MOS. Experimental results suggest that for different content videos using different encoders, the MOS would decrease with the decrease of Qp as shown in Fig.\ref{f3}\subref{f31}. Qp value is actually the serial number of quantization step ($Q_{\rm step}$) in video coding. In order to further verify the quality variable law of compressed videos, we also observed the relationship between MOS and $Q_{\rm step}$, as shown in Fig.\ref{f3}\subref{f32}. The relationship between MOS and Qp, $Q_{\rm step}$ approximately follows an exponential decay, as shown in Eq.(\ref{e1}).

\begin{equation}
	{\rm MOS}=e^{-\alpha \cdot Q_{\rm step}},
	\label{e1}
\end{equation}
where $Q_{\rm step}$ is quantization step of a compressed video, MOS is subjective score, and $\alpha$ is a parameter to be estimated. We normalize the quality of uncompressed videos to 1. Given a parameter $\alpha$, the quality scores of compressed videos can be quickly obtained, thereby reducing the cost of database construction.

Based on the exponential decay relationship, the workload of our subjective test is greatly reduced. A subset of videos can be labeled with reduced workload. In addition, the quality of the remaining videos can be inferred. To examine the feasibility of this semi-automatic rating, a series of related experiments are conducted as follows. Firstly, we randomly select 14 videos of different scenes. Each video is compressed with Qp \{22, 27, 32, 37, 42, 47, 51\}, resulting in a total of 98 compressed videos. Secondly, we obtain the subjective scores using semi-automatic labeling and subjective experiments, respectively. In the first method, subjects are asked to rate a set of compressed videos and the remaining videos are labeled using the exponential decay law of Eq.(\ref{e1}). In the second method, all subjects are asked to score all test videos to obtain the MOS values. Thirdly, we compare the correlation of the MOS values obtained by the above two methods.

\begin{figure*}[htbp] 
	\centering
	\centering
\subfloat[][] 
{
		\centering          
		\includegraphics[width=1\textwidth]{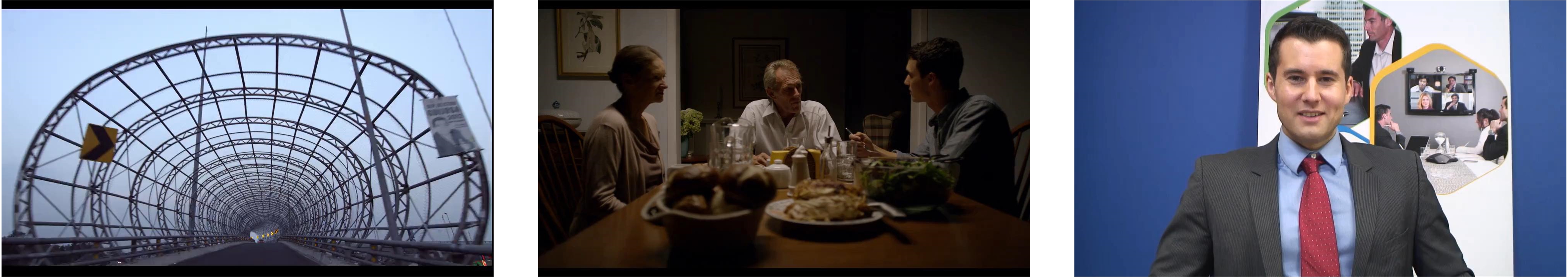}   
		\label{f41}
}

\subfloat[][] 
{
		\centering      
		\includegraphics[width=1\textwidth]{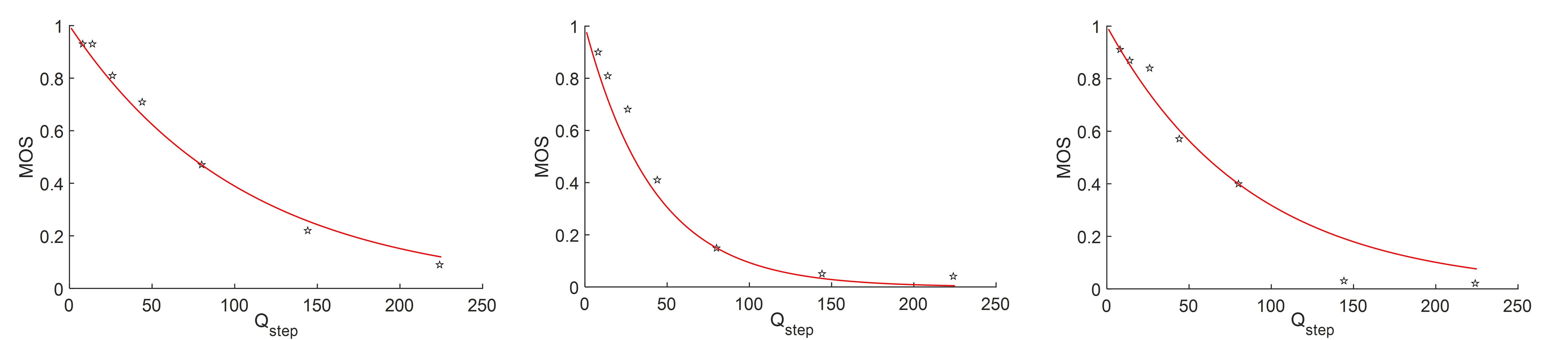}
		\label{f42}
	
}
\captionsetup{font=small,labelfont=bf}
	\caption{\label{f4}Fitting curves. (a) Compressed videos, (b) Fitted curves of $Q_{\rm step}$ and MOS.} 
\end{figure*}

Fig.\ref{f4} reveals the fitting curves of quantization step size and MOS, where the scatter points represent the MOS value, and the curve is the experimental result fitted according to Eq.(\ref{e1}). We observe that the curve basically fits the MOS values. In addition, the comparison results are presented in Table \ref{t2}, where Pearson Linear Correlation Coefficient (PLCC), Spearman Rank-order Correlation Coefficient (SRCC) and Kendall Rank Correlation Coefficient (KRCC) are utilized as performance indicators. The fitting curves and the table indicate a higher correlation between the two approaches. Clearly, semi-automatic rating is feasible for generating our large-scale video compression quality database with the advantages of low complexity and high accuracy.
	\begin{table}[ht]
	\centering\small{
		\caption{Correlation of semi-automatic labeling with MOS\label{t2}}
		\begin{tabular}{ccccc}
			\specialrule{1.5pt}{3pt}{3pt}
			\multicolumn{1}{c}{Indicators} &\multicolumn{1}{c}{PLCC} & \multicolumn{1}{c}{SRCC} &\multicolumn{1}{c}{KRCC}&\multicolumn{1}{c}{RMSE}\\
			\specialrule{0.05em}{2pt}{2pt}
			\multicolumn{1}{c}{Correlation} &\multicolumn{1}{c}{0.9815} & \multicolumn{1}{c}{0.9639} &\multicolumn{1}{c}{0.8542}&\multicolumn{1}{c}{0.0592}\\
			\specialrule{1.5pt}{2pt}{0pt}
			
	\end{tabular}}\\
\end{table}

In fact, some scholars have studied the quality change law of compressed videos. The Q-STAR \cite{44} algorithm built a video quality prediction model based on the relationship between video quality and quantization step size, temporal resolution and spatial resolution, respectively. The three parameters mean that each group of videos requires three videos to participate in subjective labeling, which is less cost-effective. The functional relationship is expressed as:
\begin{equation}
	Q=\frac{1-e^{-\alpha(\frac{s_{\rm min}}{s})}}{1-e^{-\alpha}},
	\label{e2}
\end{equation}
where {\it s} represents quantization step size. $s_{\rm min}$ is the selected minimum quantization step size. $Q$ denotes the quality of a video, and $\alpha$ is the parameter to be estimated.

In additon, Ma {\it {et al}}. \cite{45} explored the relationship between quantization step size and frame rate to predict the quality of compressed videos. The functional relationship between the video quality and the quantization step size is:
\begin{equation}
	Q=e^{c}e^{-c\frac{s}{s_{\rm min}}},
	\label{e3}
\end{equation}
where {\it c} is the parameter to be estimated.

It is worth noting that Eqs.(\ref{e1})--(\ref{e3}) are only used to reveal the relationship between the quality of video content and encoding parameters, which can only assist in building a database and cannot be directly applied in video quality evaluation tasks. In each encoder, for each video content, one of its MOS value and quantization step size must be known to determine the corresponding power exponential decay function, which can be done by subjective experiments in building the database. Video quality evaluation requires a fixed model for different video content and is not applicable. Therefore, a no-reference compressed video quality assessment method is desired to design for compressed videos. In order to determine the function building the database, the above three functions are adopted to predict the quality of videos, respectively. We randomly selected 14 videos of different scenes for verification experiments. Each video utilizes 7 different encoding parameters, and a total of 98 compressed videos are obtained, which are sufficient to support the reliability of the law. To verify the feasibility of semi-automatic labeling, we randomly select a scene video with a compression level as the benchmark to obtain the parameters to be determined by Eqs.(\ref{e1})--(\ref{e3}), respectively.  Thus, three quality scores are predicted in a semi-automatic labeling manner. Finally, the above three functions are compared, and the correlation coefficient between the semi-automatic labeling result and the MOS value is shown in Table \ref{t3}.
	\begin{table}[ht]
	\centering\small{
		\caption{Correlation of the results of different functions with the MOS\label{t3}}
		\begin{tabular}{cccc}
			\specialrule{1.5pt}{3pt}{3pt}
			\multicolumn{1}{c}{} &\multicolumn{1}{c}{PLCC} & \multicolumn{1}{c}{SRCC} &\multicolumn{1}{c}{KRCC}\\
			\specialrule{0.05em}{1pt}{2pt}
			\multicolumn{1}{c}{Equation \ref{e1}} &\multicolumn{1}{c}{\underline{0.9815}} & \multicolumn{1}{c}{\textbf{0.9639}} &\multicolumn{1}{c}{\textbf{0.8542}}\\
			\multicolumn{1}{c}{Equation \ref{e2}} &\multicolumn{1}{c}{\textbf{0.9819}} & \multicolumn{1}{c}{\underline{0.9549}} &\multicolumn{1}{c}{\underline{0.8412}}\\
			\multicolumn{1}{c}{Equation \ref{e3}} &\multicolumn{1}{c}{0.9800} & \multicolumn{1}{c}{0.9515} &\multicolumn{1}{c}{0.8257}\\
			\specialrule{1.5pt}{0pt}{0pt}
	\end{tabular}}\\
\end{table}

 Experimental results indicate that the above three functions have little correlation difference between the predicted quality scores and MOS. The performance of Eqs.(\ref{e2}) and (\ref{e3}) is lower than that of the original literatures. The reason is that only the quantization step size is considered here, and other parameters are ignored in order to reduce the cost of database construction. Although, the three expressions are similar, there are differences in the specific expressions. Among them, Eq.(\ref{e1}) has the best performance, and the expression is also the simplest, so we adopt Eq.(\ref{e1}) to develop compressed video quality database with semi-automatic ratings.

\subsection{Test video sequences}
\begin{figure}[htbp] 
	\captionsetup{font=small,labelfont=bf}
	\centering
	{\includegraphics[width=0.4\textwidth,height=0.3\textwidth]{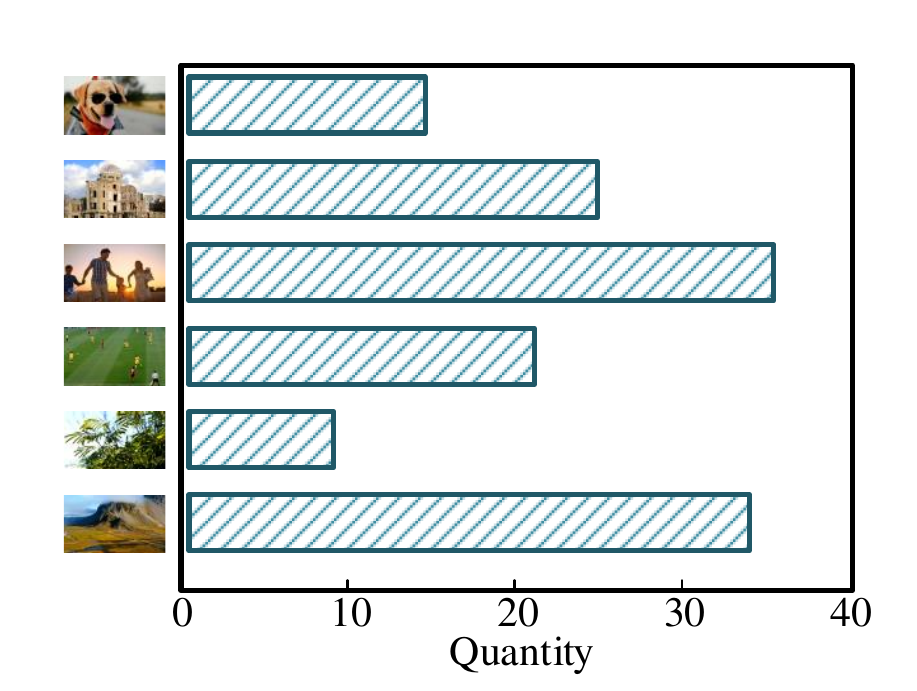}}
	\caption{Video Content Object Statistics.}
	\label{f5} 
\end{figure}

The first step in building such a large scale database is to collect a collection of high-quality and content-rich videos of natural scenes. For video content objects, we counted the number of related videos for the following 6 scenes: animals, buildings, humans, sports, plants, and landscapes. The statistical results are shown in Fig.\ref{f5}. The vertical axis refers to the scene content, and the horizontal axis represents the number of videos corresponding to each scene content. Among them, the videos describing humans are the most, and there are fewer videos of plants, but they are also close to 10. Therefore, the video contents of the proposed database are
rich and varied.

\begin{figure}[htbp] 
	\captionsetup{font=small,labelfont=bf}
	\centering
	{\includegraphics[width=0.5\textwidth,height=0.3\textwidth]{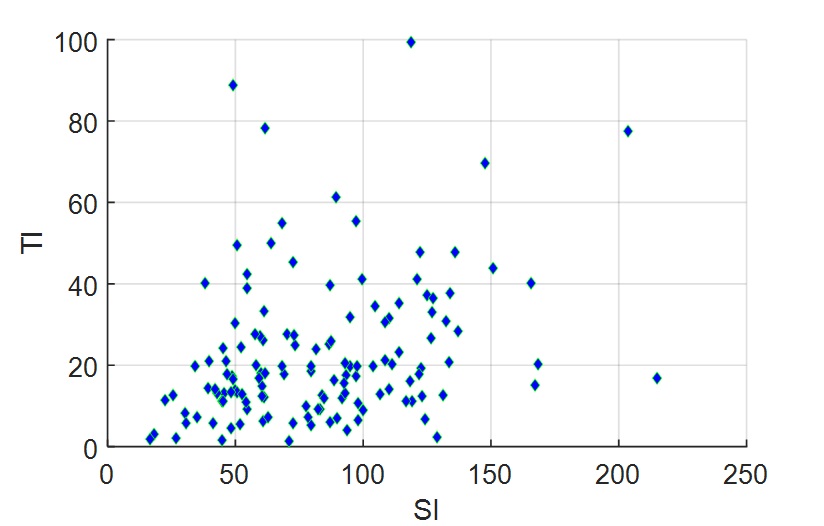}}
	\caption{The distributions of SI and TI.}
	\label{f6} 
\end{figure}

To further examine the representative of these video sequences, we also calculate their Spatial Information (SI) and Temporal Information (TI) values. The SI and TI were defined in ITU-T P. 910 \cite{47} to depict the maximal spatial gradient intensity and maximal temporal discontinuity of video contents, respectively. As shown in Fig.\ref{f6}, the horizontal axis represents SI, and the vertical axis is TI, where the maximum value of TI is close to 100, which has relatively violent motion information; the maximum value of SI exceeds 200, indicating that the video has very rich detailed information. Therefore, the selected sequences cover a vast region of SI and TI values, which are sufficiently representative and meet the requirements of the database construction. 

\subsection{Video compression method}
\begin{figure}[htbp] 
	\centering
	{\includegraphics[width=0.5\textwidth,height=0.15\textwidth]{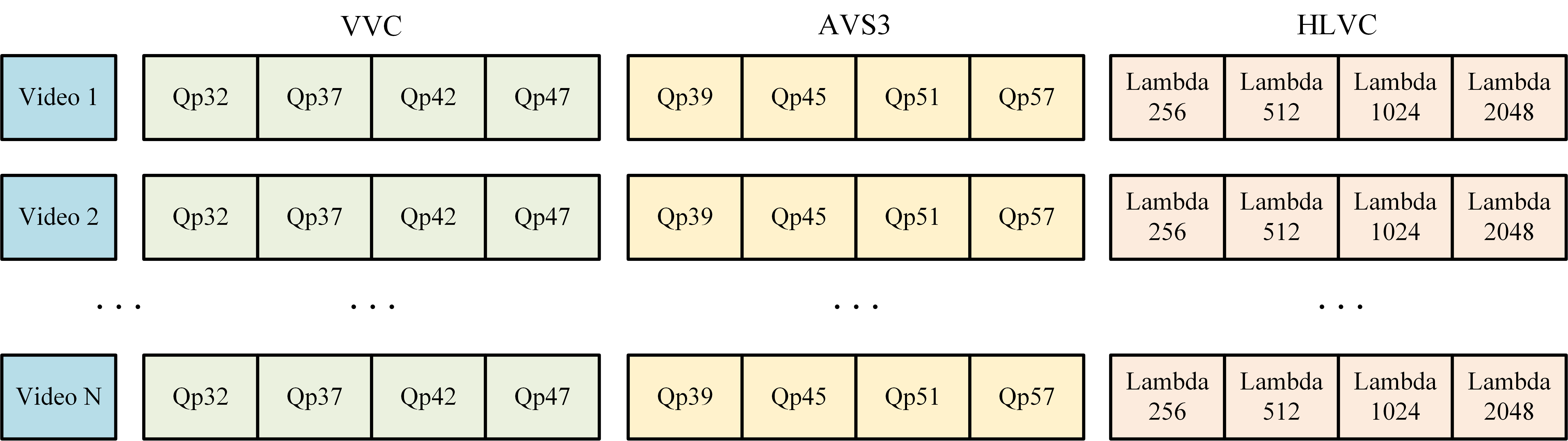}}
	\caption{Encoding information of the CVSAR database.}
	\label{f7} 
\end{figure}

The above video sequences are with the resolutions 1920$\times$1080 and 1280$\times$720, where four Qp values of 32, 37, 42 and 47 are utilized. The encoding process adopts VVC , AVS3 and HLVC \cite{31} encoders and their corresponding configuration files. It can be seen from Fig.\ref{f3} that the difference in video quality between Qp below 32 and Qp above 47 is not significant. Therefore, the Qp for VVC encoding is chosen to be the middle four, i.e., Qp=\{32, 37, 42, 47\}. According to the quantization steps corresponding to the Qp, the Qp values chosen for AVS3 compression are roughly the same as those for VVC, i.e., Qp=\{39, 45, 51, 57\}, as shown in Fig.\ref{f7}. The quality level of HLVC is mainly controlled by the hyperparameter lambda, here we adopt the four lambda =\{256, 512, 1024, 2048\} in the original text.

The encoding process utilizes official encoders and their corresponding configuration files provided by VVC and AVS3, as well as official codes published by HLVC. There are 130$\times$3$\times$4=1560 videos in total.

\subsection{Semi-automatic Labeling}
Our testing procedure follows the ITU-R BT.500 \cite{32} document with two phases. In the pre-training phase, all subjects are told about our testing procedures and trained to score videos of different quality levels. In the formal-testing phase, all subjects are asked to watch and rate 390 videos with a workload of 1.5 hours. The test sequences are presented in random order, which are displayed on a 4K screen. 23 subjects participated in the subjective experiment, including 12 males and 11 females-aged between 20 to 25. Then, we utilize the semi-automatic rating approach to calculate the subjective quality scores of the remaining 1170 videos by Eq.(\ref{e1}). By contrast, a full subjective test of all videos takes 4.5 hours per subject. Therefore, the semi-automatic rating significantly reduces the workload of subjective test but generates the inferred MOS (iMOS) values that are highly correlated to human ratings. 

\subsection{Formation of the CVSAR Database}
Based on the above methods, we construct a so far the largest of compressed video quality database CVSAR. CVSAR contains rich video content, covering various scenes such as animals, buildings, humans, sports, plants and landscapes, as shown in Table \ref{t4}, a total of 130 different high-quality videos with resolutions of 1280$\times$720 and 1920$\times$1080.

\begin{table}[ht]
	\centering{
		\caption{Video content distribution of the CVSAR database\label{t4}}
		\newcommand{\tabincell}[2]{\begin{tabular}{@{}#1@{}}#2\end{tabular}}
		\begin{tabular}{ccc}
			\specialrule{1.25pt}{3pt}{3pt}
			\multicolumn{1}{c}{Video content} &\tabincell{cc}{Number of\\original videos} &\tabincell{c}{Number of\\compressed videos} \\
			\specialrule{0.05em}{2pt}{2pt}
			\multicolumn{1}{c}{Animal}&\multicolumn{1}{c}{13}&\multicolumn{1}{c}{156}\\
			\multicolumn{1}{c}{Architecture}&\multicolumn{1}{c}{24}&\multicolumn{1}{c}{288}\\
			\multicolumn{1}{c}{Human}&\multicolumn{1}{c}{34}&\multicolumn{1}{c}{408}\\
			\multicolumn{1}{c}{Motion}&\multicolumn{1}{c}{21}&\multicolumn{1}{c}{252}\\
			\multicolumn{1}{c}{Plant}&\multicolumn{1}{c}{9}&\multicolumn{1}{c}{108}\\
			\multicolumn{1}{c}{Scenery}&\multicolumn{1}{c}{29}&\multicolumn{1}{c}{348}\\
			\specialrule{1.25pt}{0pt}{0pt}
	\end{tabular}}\\
\end{table}

In total, there are 1560 compressed videos in the CVSAR database with different contents, resolutions, encoders and Qps, which are presented in Table \ref{t5}.
\begin{table}[ht]
	\centering{
		\caption{Overview of the CVSAR database\label{t5}}
		\begin{tabular}{cc}
			\specialrule{1.25pt}{3pt}{3pt}
			\multicolumn{1}{c}{Database information} &\multicolumn{1}{c}{Specific settings}\\
			\specialrule{0.05em}{2pt}{2pt}
			\multicolumn{1}{c}{Encoder} &\multicolumn{1}{c}{VVC, AVS3, HLVC}\\
			\multicolumn{1}{c}{Resolution} &\multicolumn{1}{c}{1280$\times$720,1920$\times$1080}\\
			\multicolumn{1}{c}{Qp of VVC} &\multicolumn{1}{c}{\{32, 37, 42, 47\}}\\
			\multicolumn{1}{c}{Qp of AVS3} &\multicolumn{1}{c}{\{39, 45, 51, 57\}}\\
			\multicolumn{1}{c}{Lambda of HLVC} &\multicolumn{1}{c}{\{256, 512, 1024, 2048\}}\\
			\multicolumn{1}{c}{Number of videos} &\multicolumn{1}{c}{960  videos with 720p, 600 videos with 1080p }\\
			\specialrule{1.25pt}{0pt}{0pt}
	\end{tabular}}\\
\end{table}

\section{Proposed STFEE model}
We propose an end-to-end no-reference quality assessment method for compressed videos. The proposed model, namely STFEE, is a 3D CNN to predict the perceptual quality of compressed videos. The network architecture is illustrated in Fig.\ref{f9}. Firstly, the input video sequence is equally divided into several sub-sequences of the same size, and each sub-sequence is divided into multiple small cubes of the same size in a non-overlapping manner. Then, the video sub-sequences are fed into our 3D CNN to extract spatiotemporal features, which are utilized as the input of the transformer regression network. Finally,  the regression network performs long-term memory-dependent learning on the spatiotemporal features of sub-sequences in different time periods and extracts the global features of the corresponding videos. The global features are regressed onto the quality score of compressed videos by fully connected layers. In the following subsections, video preprocessing, spatiotemporal features extracting, spatiotemporal features regressing and model training will be discussed in detail.

\subsection{Video preprocessing}
Visual saliency is an inherent attribute of Human Visual System (HVS) and is also a key factor affecting video perceptual quality \cite{48}. The advantages of introducing visual saliency into video quality assessment are primarily reflected in two aspects: first, it allocates constrained hardware resources to more significant regions; second, video quality analysis considering visual saliency is more consistent with human visual perception. Therefore, we select improved HED \cite{34} as our video saliency model based on comprehensive comparison and analysis of popular video saliency models. The saliency model has strong applicability and its high accuracy. 

\begin{figure*}[htbp!] 
	\captionsetup{font=small,labelfont=bf}
	\centering
	{\includegraphics[width=18cm,keepaspectratio]{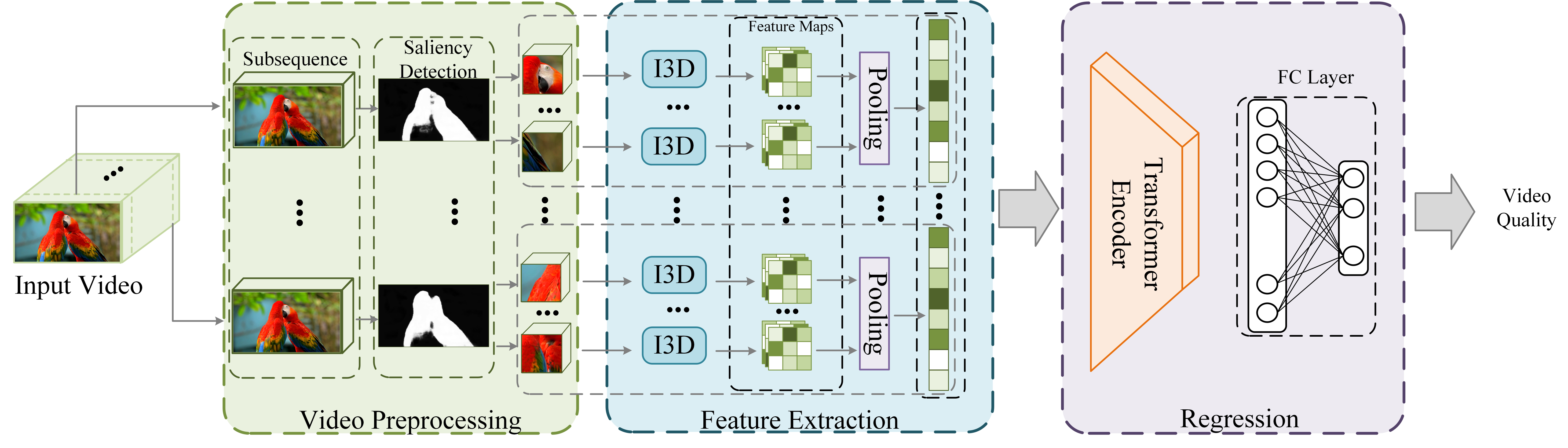}}
	\caption{Proposed STFEE Framework.}
	\label{f9} 
\end{figure*}

Based on the saliency model, video preprocessing process is shown in Fig.\ref{f9}. First, the subsequences of videos are cut every half second, and each subsequence is a continuous 16 frames. Then, salient regions are extracted by using a saliency detection algorithm. Finally, video segmentation is performed based on the minimum circumscribed rectangle. The size of video blocks is set to 224$\times$224. Small cubes are cut along the timeline. The dimension of each cube is 224$\times$224$\times$3$\times$16, where 3 and 16 are the number of channels and consecutive frames, respectively.
\subsection{Spatiotemporal feature extraction}
3D convolution kernel can effectively extract video spatiotemporal features. Therefore, we select I3D \cite{35} as our video feature extraction model based on comprehensive comparison and analysis of popular 3D convolutional networks. Based on the I3D network, the spatiotemporal features of each small cube are extracted. The features of sub-sequences are obtained by pooling the spatiotemporal features of all its small cubes. Since the 5$\times$5$\times$5 convolution kernel of I3D network causes a large amount of computation, we utilize a 1$\times$1$\times$1 convolution kernel for dimensionality reduction. Furthermore, attention mechanism is able to optimize feature extraction during network training. In order to enhance the learning ability of the network, an attention mechanism is introduced into the convolutional layers of the last two Inception Modules. On the basis of comprehensive consideration of feature extraction and computational complexity, a channel attention module is introduced into the convolutional layers of the last two Inception Modules of the network structure.

Based on our improved I3D network, each small cube will obtain a 1$\times$1024 dimensional feature ,which is as follows:
\begin{equation}
	\begin{aligned}
	&F_{\rm seq}={\rm Pool}(F_i), i\in{1,2,...,N_{\rm cube}},\\
	&F={\rm I3D}({\rm cube}),
	\label{e8}
	\end{aligned}
\end{equation}
where cube represents a small cube. I3D() denotes feature extraction operation. {\it F} is spatiotemporal feature of each small cube. Pool\{\} refers to a pooling operation, which pools the spatiotemporal features of all small cubes in a subsequence into subsequence-level spatiotemporal features, namely $F_{\rm seq}$. $N_{\rm cube}$ refers to the number of small cubes divided by each subsequence.

In the same sub-sequence, the feature vectors of several small cubes are combined into a feature map {\it F} with dimension {\it N}$\times$1024, where {\it N} indicates that there are {\it N} small cubes in the subsequence. The pooling operation is performed on the feature map {\it F} to obtain spatiotemporal features of the sub-sequences.
\begin{equation}
	\begin{aligned}
		&\textbf{Feature Pooling:}\\
		&F_{\rm avg}={\rm AvgPool}(F); \;{\rm shape}=(1;1024);\\
		&F_{\rm max}={\rm MaxPool}(F); \;{\rm shape}=(1;1024);\\
		&F_{\rm seq}=(F_{\rm avg};F_{\rm max}); \; {\rm shape}=(1;2048);\\
	\end{aligned}
\label{e4}
\end{equation}

Among them, $F_{\rm avg}$ and $F_{\rm max}$ are the results of performing mean pooling and maximum pooling according to the first dimension of the feature map {\it F}, respectively. Then, these two feature vectors are directly concatenated into a more representative feature vector $F_{\rm seq}$. For a 5-second video, 16 frames of video segments are intercepted every half second to obtain 10 sub-sequences, thus generating 10 spatiotemporal features $F_{\rm seq}$, which constitute the global spatiotemporal feature ${\rm Global}_F$ of videos, with a dimension of 10$\times$2048.
\subsection{Model Learning}
For an input compressed video, the proposed STFEE network {\it M} is utilized to predict the perceptual quality  $Q_{\rm pred}$ of compressed videos:
\begin{equation}
	Q_{\rm pred}=M(V,\theta),
	\label{e5}
\end{equation}
where $\theta$ indicates all parameters of this network.

Denote the ground truth quality of the input as ${\rm QC}_{\rm label}$ .The training goal of network {\it M} is to find the optimal parameter setting, so as to minimize the overall quality prediction loss between ${\rm QC}_{\rm pred}$ and ${\rm QC}_{\rm label}$ of all video cubes in the training dataset. We apply the MSE as loss function in the training process, which is widely used in various regression tasks.
\begin{equation}
	{\rm Loss}_{\rm cube}=\frac{1}{N}\sum_{i=1}^N{\Vert {\rm QC}_{\rm pred}(i)-{\rm QC}_{\rm label}(i) \Vert_2},
	\label{e6}
\end{equation}
where ${\rm QC}_{\rm pred}(i)$ and ${\rm QC}_{\rm label}(i)$ refer to the predicted quality and MOS value of the {\it i}-th cube, respectively. {\it N} represents the number of input cubes. The SGD algorithm is utilized with a learning rate of 0.001.

The spatiotemporal features of each video subsequence are obtained according to Eq.(\ref{e4}). Further, the global spatiotemporal feature ${\rm Global}_F$ of the entire video can be obtained. In order to learn its long-term dependency information, it is used as the input of the Transformer Encoder. Then, the overall quality score of the video is predicted through the fully connected layer regression. The loss function is:
\begin{equation}
    {\rm Loss}=\frac{1}{K}\sum_{i=1}^K{\Vert  Q_{\rm pred}(i)-Q_{\rm label}(i) \Vert_1},
	\label{e7}
\end{equation}
where $Q_{\rm pred}(i)$ and $Q_{\rm label}(i)$ denote the predicted quality and ground truth of the {\it i}-th video, respectively. {\it K} is the number of input videos. The Adam algorithm is utilized with a learning rate of 0.001.
\section{EXPERIMENTAL RESULTS}
In this section, the performance of the proposed STFEE model is evaluated and compared with typical video quality metrics on four datasets. In addition, we analyze the effect of block sizes on the performance of spatiotemporal features. Finally, the cross-database test is also performed to further verify the generalization performance of the algorithm.

\begin{table*}[!htbp] 
	\centering
	\caption{PERFORMANCE COMPARISON OF STFEE METHODS\label{t6}}
	\begin{tabular}{ccc|cc|cc|cc} 
		\specialrule{1.25pt}{3pt}{3pt}
		\multirow{2}{0.2\textwidth}{Methods}& \multicolumn{2}{c}{CVSAR} & \multicolumn{2}{c}{LIVE} &\multicolumn{2}{c}{CSIQ} &\multicolumn{2}{c}{WaterlooIVC4K}\\
		\cmidrule(r){2-3}\cmidrule(r){4-5}\cmidrule(r){6-7}\cmidrule(r){8-9}
		&PLCC & SRCC &PLCC & SRCC &PLCC & SRCC&PLCC & SRCC\\
		\specialrule{0.75pt}{2pt}{2pt}
		\multicolumn{1}{l}{PSNR}     &0.2964&0.3471    &0.5122&0.4790       &0.5442&0.5651    &0.3049&0.3097\\   
		\multicolumn{1}{l}{SSIM}     &0.4606&0.5673    &0.5405&0.5863       &0.5318&0.6002    &0.4290&0.4022\\
		\multicolumn{1}{l}{MS-SSIM}  &0.4734&0.6991    &0.5912&0.6485       &0.6119&0.7347    &0.5938&0.5316\\
		\multicolumn{1}{l}{SpEED-VQA}&0.3630&0.6252    &0.6404&\textbf{0.7681}&\underline{0.7371}&\underline{0.7413}   &0.5328&0.4739\\
		\multicolumn{1}{l}{NIQE}     &0.4358&0.4205    &0.2917&0.1178       &0.4428&0.4282   &0.1182&0.2467\\
		\multicolumn{1}{l}{VIIDEO}   &0.2617&0.1829    &0.6380&0.6057       &0.3087&0.1234   &0.0094&0.0107\\
		\multicolumn{1}{l}{3D-PSD}   &0.7008&0.7042    &0.1737&0.1633       &0.6321&0.5525   &0.5922&0.5106\\
		\multicolumn{1}{l}{TLVQM}   &0.6923&0.6845     &0.3795&0.3673       &0.6239&0.6178   &0.7586&0.7627\\
		\multicolumn{1}{l}{VSFA}    &0.7746&0.7687&0.4126&0.6102&0.6752&0.7324 &0.4963&0.5056\\
		\multicolumn{1}{l}{MDTVSFA} &0.7781&0.7619     &0.4136&0.5404       &0.5519&0.5820   &\textbf{0.8937}&\textbf{0.8927}\\
		\multicolumn{1}{l}{GSTVQA}  &0.7795&0.7444&0.7150&0.7237&0.5725&0.5806&0.4567&0.4391\\
		\multicolumn{1}{l}{Shen's algorithm} &\underline{0.7813}&\underline{0.7932}&\underline{0.7182}&0.6463&0.6790&0.7369&0.4101&0.3496\\
		\specialrule{0.5pt}{2pt}{2pt}
		\multicolumn{1}{l}{STFEE}&\textbf{0.9203}&\textbf{0.9098}&\textbf{0.7352}&\underline{0.7451}&\textbf{0.7883}&\textbf{0.7682}&\underline{0.8813}&\underline{0.8775}\\
		\specialrule{1.25pt}{3pt}{3pt}
	\end{tabular}
\end{table*}

\subsection{Experimental setups}
We train the STFEE model on the proposed CVSAR database. To verify the generality of the proposed method, we also select three publicly available databases, LIVE, CSIQ, WaterlooIVC4K, for cross-database validations. 

CVSAR contains 1560 compressed videos, which are split into 80:20 training/testing sets according to the contents. In addition, we utilize 5-fold cross-validation to evaluate the performance of STFEE. The performance shown in Table \ref{t8} is the average performance across all test sets.

LIVE is a video quality evaluation database containing 150 distorted videos. It contains 10 reference videos with a resolution of 680$\times$432. Four types of distortions are presented with video data. Among them, the distortion of compression is generated by H.264 and MPEG-2 encoder, resulting 80 outputs. This subset is utilized here to evaluate the performance of STFEE.   

CSIQ is a collection of 216 distorted videos. It contains 12 reference videos. Three types of distortions are presented with video data, which are H.264, HEVC and MJPEG. Each compression corresponds to 3 levels, resulting 108 outputs. 

WaterlooIVC4K is created from 20 reference videos with a resolution of 3840$\times$2160. All videos are collected from Youtube Creative Commons videos. Each source video is encoded by five encoders: HEVC, H.264, VP9, AV1 and AVS2. Each source video is encoded in three resolutions: 960$\times$540, 1920$\times$1080, 840$\times$2160. Each setting has four distortion levels, where the encoder control parameters are determined to ensure good perceptual separation. Finally, a total of 1200 encoded videos are produced.

There is no uniform score range and type in these databases. The setups of CVSAR are chosen as our standard in these experiments. Subjective scores of the other three databases are normalized to the range of [0,1].

\subsection{Performance Comparison}
(1) Performance comparison on different databases

To verify the performance of the proposed STFEE, it is evaluated on the CVSAR, LIVE, CSIQ and WaterlooIVC4KV databases. The method of STFEE is also compared with typical video quality metrics including PSNR, SSIM, MS-SSIM \cite{36}, 3D-PSD \cite{37}, SpEED-VQA \cite{7}, NIQE \cite{39}, VIIDEO \cite{23}, TLVQM \cite{40}, VSFA \cite{12}, MDTVSFA \cite{41}, GSTVQA \cite{13} and Shen's algorithm \cite{46} to show its performance. We obtain the source codes of these metrics from the author’s public websites. In addition, we utilize the PLCC and SRCC as the performance indicators. The results summarized in Table \ref{t6}, where the best and the second-best results are shown in bold and underlined, respectively. Experimental results have indicated the superior performance of our proposed STFEE. FR-VQA method focuses on comparing the differences between reference videos and compressed videos. In video encoding process, PSNR and SSIM metrics are often utilized to measure the quality of compressed videos. However, the correlation between these two algorithms and subjective perceived quality is relatively low. The reason is that, first of all, these two indicators are essentially image quality evaluations, and do not consider video motion characteristics in temporal domain. Moreover, these two indicators simply calculate the difference between image pixels and structures, without considering human perceptual characteristics. Therefore, they perform poorly in video quality evaluation. MS-SSIM is an improved SSIM method, but its performance is still not good enough. From Table VI, the following observations can be drawn:

Firstly, STFEE obtains the highest PLCC and SRCC on the CVSAR database. VSFA, MDTVSFA, GSTVQA and Shen's algorithm \cite{46} are also based on deep learning, which shows that they have better performance than manual feature extraction.

Secondly, compared to the CVSAR database, the LIVE database has only one resolution of 768$\times$432. The proposed STFEE is still competitive, where the PLCC still remains the highest. As a natural scene statistics method, NIQE is still not suitable for LIVE database. Neither 3DPSD nor TLVQM can show ideal performance on the LIVE database.

Thirdly, STFEE still exhibits the best performance on the CSIQ database. As a reduced-reference quality evaluation algorithm, SpEED-VQA has excellent performance and is more suitable for the LIVE and CSIQ databases. VSFA and Shen's algorithm \cite{46} also have good generalization performance on the CSIQ database.

Finally, WaterlooIVC4K has more videos and higher resolution than the LIVE and CSIQ databases. Except TLVQM and MDTVSFA, the performance of several other quality evaluation algorithms has decreased. The proposed STFEE still exhibits superior performance.

\begin{figure}[htbp!] 
	\centering
	{\includegraphics[width=9cm,keepaspectratio]{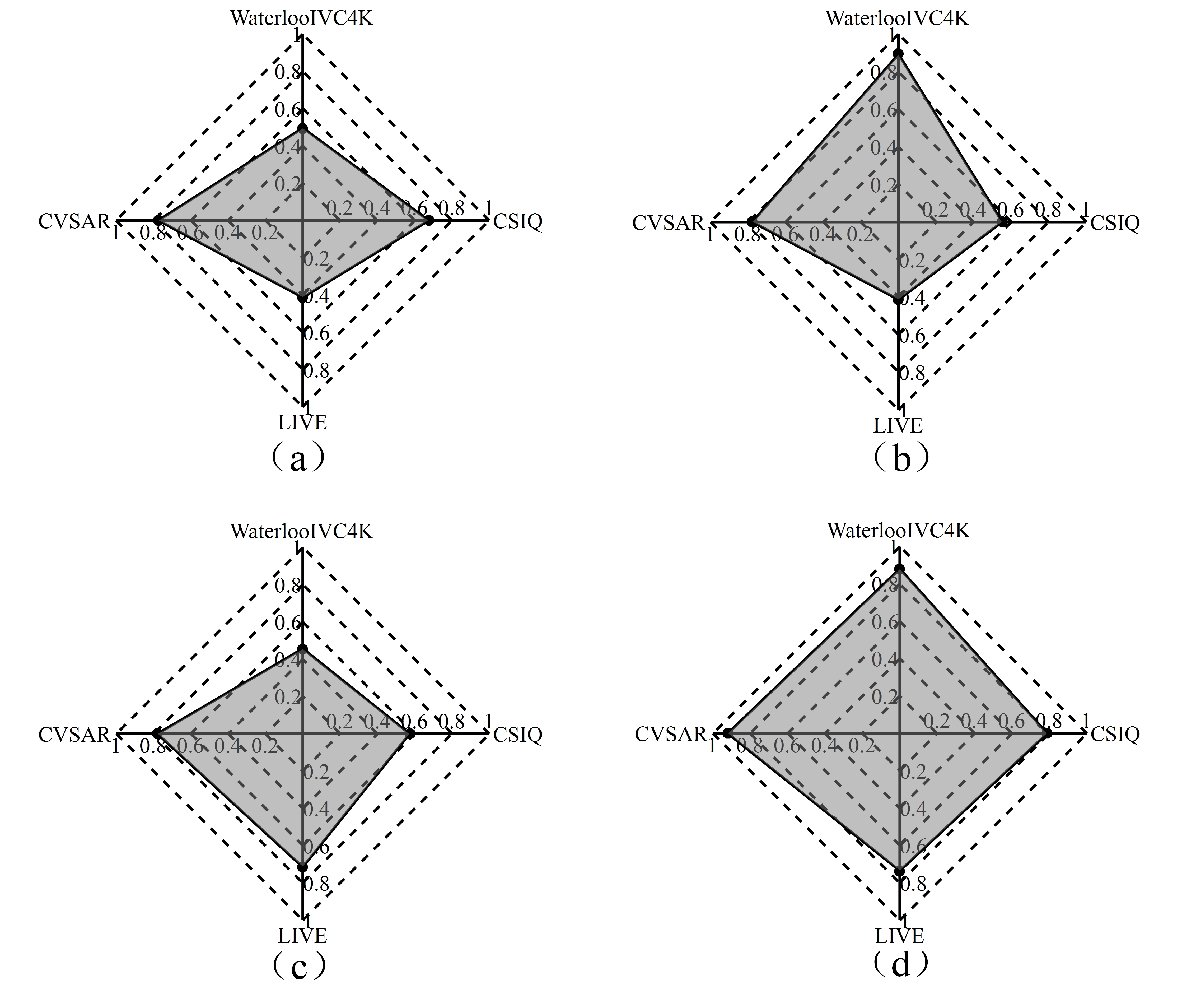}}
	\caption{PLCC Radar Chart: (a)VSFA, (b)MDTVSFA, (c)GSTVQA, (d)STFEE.}
	\label{f10} 
\end{figure}

To further verify the generalization performance of the proposed algorithm with other deep learning algorithms more intuitively, we plot the PLCC values of the VSFA, MDTVSFA, GSTVQA and STFEE algorithms on the four databases into a radar chart, as shown in Fig.\ref{f10}.

In Fig.\ref{f10}, each radar chart corresponds to an algorithm, and there are 4 coordinate axes, each of which represents a database. Fig.\ref{f10}(a) shows that the VSFA algorithm performs poorly on the LIVE and WaterlooIVC4K databases. Fig.\ref{f10}(b) illustrates that the MDTVSFA algorithm performs well on the WaterlooIVC4K database, but performs poorly on the LIVE and CSIQ databases. Fig.\ref{f10}(c) implies that GSTVQA does not perform well on the WaterlooIVC4K database. Fig.\ref{f10}(d) proves that the proposed algorithm achieves superior performance on all four databases and has good generalization performance. In addition, Fig.\ref{f10} also indicates that the large-scale database CVSAR can provide stable data support for deep learning methods, and the four algorithms all have good performance on the CVSAR database. Although the WaterlooIVC4K database also contains thousands of videos, it only contains 20 different scenarios. The CVSAR database includes 130 video contents and has good generalization performance.

(2) Video block size selection

In order to verify the effect of video block size on video spatiotemporal feature extraction, we 
perform the following experiments, as shown in Table \ref{t7}.
\begin{table*}[ht]
	\caption{STFEE  performance  on different databases\label{t8}}
	\renewcommand\arraystretch{1.25}
	\centering\small{
		\setlength\tabcolsep{8pt} 
		\begin{tabular}{cccccccccc}
			
			\specialrule{1.5pt}{3pt}{3pt}
			\multicolumn{2}{c}{\diagbox[dir=NW]{training library}{test library}} &\multicolumn{1}{c}{} & \multicolumn{2}{c}{LIVE} & \multicolumn{2}{c}{CSIQ}& \multicolumn{1}{c}{WaterlooIVC4K}& \multicolumn{2}{c}{CVSAR}\\
			
			\specialrule{1.25pt}{3pt}{3pt}
			\multicolumn{2}{c}{\multirow{2}*{LIVE}} &\multicolumn{1}{c}{PLCC} & \multicolumn{2}{c}{-} &\multicolumn{2}{c}{0.6707}&\multicolumn{1}{c}{0.4577}&\multicolumn{2}{c}{0.6037}\\
			\cline{3-10}
			\multicolumn{2}{c}{~}&\multicolumn{1}{c}{SRCC} & \multicolumn{2}{c}{-} &\multicolumn{2}{c}{0.7080}&\multicolumn{1}{c}{0.4014}&\multicolumn{2}{c}{0.6500}\\
			
			\specialrule{1.25pt}{3pt}{3pt}
			\multicolumn{2}{c}{\multirow{2}*{CSIQ}} &\multicolumn{1}{c}{PLCC} & \multicolumn{2}{c}{0.4758} &\multicolumn{2}{c}{-}&\multicolumn{1}{c}{0.4260}&\multicolumn{2}{c}{0.5827}\\
			\cline{3-10}
			\multicolumn{2}{c}{~}&\multicolumn{1}{c}{SRCC} & \multicolumn{2}{c}{0.4671} &\multicolumn{2}{c}{-}&\multicolumn{1}{c}{0.3428}&\multicolumn{2}{c}{0.5865}\\
			
			\specialrule{1.25pt}{3pt}{3pt}
			\multicolumn{2}{c}{\multirow{2}*{WaterlooIVC4K}} &\multicolumn{1}{c}{PLCC} & \multicolumn{2}{c}{0.4346} &\multicolumn{2}{c}{0.6868}&\multicolumn{1}{c}{-}&\multicolumn{2}{c}{0.7734}\\
			\cline{3-10}
			\multicolumn{2}{c}{~}&\multicolumn{1}{c}{SRCC} & \multicolumn{2}{c}{0.4449} &\multicolumn{2}{c}{0.7042}&\multicolumn{1}{c}{-}&\multicolumn{2}{c}{0.7525}\\
			\specialrule{1.25pt}{3pt}{3pt}
			
			\multicolumn{2}{c}{\multirow{2}*{CVSAR}} &\multicolumn{1}{c}{PLCC} & \multicolumn{2}{c}{0.3473} &\multicolumn{2}{c}{0.7051}&\multicolumn{1}{c}{0.6556}&\multicolumn{2}{c}{-}\\
			\cline{3-10}
			\multicolumn{2}{c}{~}&\multicolumn{1}{c}{SRCC} & \multicolumn{2}{c}{0.4459} &\multicolumn{2}{c}{0.7241}&\multicolumn{1}{c}{0.6071}&\multicolumn{2}{c}{-}\\
			\specialrule{1.5pt}{3pt}{3pt}
	\end{tabular}}
\end{table*}

\begin{table}[ht]
	\centering\small{
		\caption{Network performance of different video block sizes\label{t7}}
		\begin{tabular}{cccc}
			\specialrule{1.25pt}{3pt}{3pt}
			\multicolumn{1}{c}{size} &\multicolumn{1}{c}{PLCC} & \multicolumn{1}{c}{SRCC} &\multicolumn{1}{c}{KRCC}\\
			\specialrule{0.05em}{2pt}{2pt}
			\multicolumn{1}{c}{64$\times$64} &\multicolumn{1}{c}{0.4285} & \multicolumn{1}{c}{0.4124} &\multicolumn{1}{c}{0.2823}\\
			\multicolumn{1}{c}{128$\times$128} &\multicolumn{1}{c}{0.6038} & \multicolumn{1}{c}{0.5806} &\multicolumn{1}{c}{0.4087}\\
			\multicolumn{1}{c}{224$\times$224} &\multicolumn{1}{c}{\textbf{0.7391}} & \multicolumn{1}{c}{\textbf{0.7322}} &\multicolumn{1}{c}{\textbf{0.5387}}\\
			\multicolumn{1}{c}{256$\times$256} &\multicolumn{1}{c}{\underline{0.6842}} & \multicolumn{1}{c}{\underline{0.6707}} &\multicolumn{1}{c}{\underline{0.4819}}\\
			\specialrule{1.25pt}{0pt}{0pt}
	\end{tabular}}\\
\end{table}

Experimental results indicate that when the size of video cubes is 224$\times$224, its network performance is the best. The resolution of 256$\times$256 reduces the learning ability of the network, and requires higher computing power of hardwares. Therefore, we choose 224$\times$224 as the size of small cubes, which enables the network to learn more effective spatiotemporal features.

(3) Validation across datasets

To further verify the generalization performance of the proposed STFEE, cross-dataset validation is performed. We utilize LIVE, CSIQ, WaterlooIVC4K and our proposed CVSAR database as training sets to obtain network models. Then, the other three datasets are used as test sets to test their performance, as shown in Table \ref{t8}.

The above experimental results show that the proposed STFEE has good generalization ability on different databases. Among them, the performance differences on different databases are mainly due to different database scales and different compression methods. The LIVE and CSIQ databases contain a small number of compressed videos, so the models trained on these two databases are not expressive enough. And for the larger waterlooIVC4K and CVSAR datasets, the performance of the algorithm is relatively better. The above experiments also suggest that the size of the dataset is a key factor affecting the network performance.

\section{CONCLUSIONS}
In this work, we exploit the exponentially decaying relationship between quantization step size and compressed video quality, and propose a semi-automatic rating method that greatly reduces the labeling workload while maintaining high labeling accuracy. Utilizing this approach, we build the CVSAR database, which is currently the largest database for compressed videos and has the richest scene content. Then, we develop an end-to-end STFEE model for compressed videos, which adopts a 3D convolutional network for feature extraction and follows by a transformer for quality regression. By training on the CVSAR database, the STFEE model outperforms the state-of-the-art VQA algorithms. Cross-database validation also reveals the generalization ability of our STFEE model. We will make the proposed database and quality model publicly available to facilitate reproducible research.

\end{sloppypar}

\begin{thebibliography}{99}    
	
	\bibitem{1} W. Liu, Z. Duanmu and Z. Wang, “End-to-end blind quality assessment of compressed videos using deep neural networks,” in {\it Proc. 2018 ACM Multimedia Conference (ACM MM)}, 2018, pp. 546–554.
	\bibitem{2} A. C. Bovik, “Automatic Prediction of Perceptual Image and Video Quality,” {\it Proceedings of the IEEE}, vol. 101, no. 9, pp. 2008-2024, Sept. 2013.
	\bibitem{3} A. Horé and D. Ziou, “Image Quality Metrics: PSNR vs. SSIM,” in {\it Proc. 20th Int. Conf. Pattern Recognition (ICPR)}, 2010, pp. 2366-2369.
	\bibitem{4} Z. Wang, A. C. Bovik, H. R. Sheikh and E. P. Simoncelli, “Image quality assessment: from error visibility to structural similarity,” {\it IEEE Trans. Image Process.}, vol. 13, no. 4, pp. 600-612, Apr. 2004.
	\bibitem{5} H. Fu, D. Pan and P. Shi, “Full-Reference Video Quality Assessment Based on Spatiotemporal Visual Sensitivity,” in {\it Proc. Int. Conf. Culture-oriented Science \& Technology (ICCST)}, 2021, pp. 305-309.
	\bibitem{6} T. Li, X. Min, H. Zhao, G. Zhai, Y. Xu and W. Zhang, “Subjective and Objective Quality Assessment of Compressed Screen Content Videos,” {\it IEEE Trans. Broadcast.}, vol. 67, no. 2, pp. 438-449, Jun. 2021.
	\bibitem{7} C. G. Bampis, P. Gupta, R. Soundararajan and A. C. Bovik, “SpEED-QA: Spatial Efficient Entropic Differencing for Image and Video Quality,” {\it IEEE Trans. Signal Process. Lett.}, vol. 24, no. 9, pp. 1333-1337, Sept. 2017.
	\bibitem{8} J. Korhonen, “Study of the Subjective Visibility of Packet Loss Artifacts in Decoded Video Sequences,” {\it IEEE Trans. Broadcast.}, vol. 64, no. 2, pp. 354-366, Jun. 2018.
	\bibitem{9} M. A. Usman, M. R. Usman and S. Y. Shin, “A Novel No-Reference Metric for Estimating the Impact of Frame Freezing Artifacts on Perceptual Quality of Streamed Videos,” {\it IEEE Trans. Multimedia}, vol. 20, no. 9, pp. 2344-2359, Sept. 2018.
	\bibitem{10} S. V. Reddy Dendi and S. S. Channappayya, “No-Reference Video Quality Assessment Using Natural Spatiotemporal Scene Statistics,” {\it IEEE Trans. Image Process.}, vol. 29, pp. 5612-5624, 2020.
	\bibitem{11} A. Rehman, K. Zeng, and Z. Wang, “Display device-adapted video quality-of-experience assessment,” in {\it Proc. SPIE 9394, Human Vision and Electronic Imaging XX}, 2015, pp. 939406.
	\bibitem{12} D. Li, T. Jiang, M. Jiang, “Quality assessment of in-the-wild videos,” in {\it Proc. 27th ACM Int. Conf. Multimedia (ACM MM)}, 2019, pp. 2351-2359.
	\bibitem{13} B. Chen, L. Zhu, G. Li, F. Lu, H. Fan and S. Wang, “Learning Generalized Spatial-Temporal Deep Feature Representation for No-Reference Video Quality Assessment,” {\it IEEE Trans. Circuits Syst. Video Technol.}, vol. 32, no. 4, pp. 1903-1916, Apr. 2022.
	\bibitem{14} Y. Zhang, H. Liu, Y. Yang, X. Fan, S. Kwong and C. C. J. Kuo, “Deep Learning Based Just Noticeable Difference and Perceptual Quality Prediction Models for Compressed Video,” {\it IEEE Trans. Circuits Syst. Video Technol.}, vol. 32, no. 3, pp. 1197-1212, Mar. 2022.
	\bibitem{15} K. Seshadrinathan, R. Soundararajan, A. C. Bovik and L. K. Cormack, “Study of Subjective and Objective Quality Assessment of Video,” {\it IEEE Trans. Image Process.}, vol. 19, no. 6, pp. 1427-1441, Jun. 2010.
	\bibitem{16} A. K. Moorthy,  L. K. Choi, G. De Veciana and A. Bovik, “Mobile video quality assessment database,” {\it IEEE ICC Workshop on Realizing Advanced Video Optimized Wireless Networks}, 2012, vol. 6, no. 6, pp. 652-671.
	\bibitem{17} P. V. Vu and D. M. Chandler, “ViS3: An algorithm for video quality assessment via analysis of spatial and spatiotemporal slices,” {\it Journal of Electronic Imaging}, vol. 23, no. 1, pp. 013016, 2014.
	\bibitem{18} T. Blu, W. Cham, K. Ngan, {\it et al.}, “IVP video quality database,” [Online]. Available:http://ivp.ee.cuhk.edu.hk/, 2018.
	\bibitem{19} Z. Li, Z. Duanmu, W. Liu, {\it et al.}, “AVC, HEVC, VP9, AVS2 or AV1? — A comparative study of state-of-the-art video encoders on 4K videos,” in {\it Proc. 16th Int. Conf. Image Anal. and Recognit. (ICIAR)}, 2019, pp. 27-29.
	\bibitem{20} L. Choi and A. Bovik, “Video quality assessment accounting for temporal visual masking of local flicker,” {\it Signal Process. Image Commun.}, vol. 67, pp. 182-198, 2018. 
	\bibitem{21}J. Korhonen, “Study of the Subjective Visibility of Packet Loss Artifacts in Decoded Video Sequences,” {\it IEEE Trans. Broadcast.}, vol. 64, no. 2, pp. 354-366, Jun. 2018.
	\bibitem{22} M. A. Saad, A. C. Bovik and C. Charrier, “Blind Prediction of Natural Video Quality,” {\it IEEE Trans. Image Process.}, vol. 23, no. 3, pp. 1352-1365, Mar. 2014.
	\bibitem{23} A. Mittal, M. A. Saad and A. C. Bovik, “A Completely Blind Video Integrity Oracle,” {\it IEEE Trans. Image Process.}, vol. 25, no. 1, pp. 289-300, Jan. 2016.
	\bibitem{24} S. V. Reddy Dendi and S. S. Channappayya, “No-Reference Video Quality Assessment Using Natural Spatiotemporal Scene Statistics,” {\it IEEE Trans. Image Process.}, vol. 29, pp. 5612-5624, Apr. 2020.
	\bibitem{25} Y. Zhang, H. Liu, Y. Yang, X. Fan, S. Kwong and C. C. J. Kuo, “Deep Learning Based Just Noticeable Difference and Perceptual Quality Prediction Models for Compressed Video,” {\it IEEE Trans. Circuits Syst. Video Technol.}, vol. 32, no. 3, pp. 1197-1212, Mar. 2022.
	\bibitem{26} A. Vaswani, N. Shazeer, N. Parmar {\it et al.}, “Attention is all you need,” in {\it Neural Information Processing Systems (NIPS)}, 2017, pp. 5998-6008. 


	 
	


	\bibitem{43} T. Zhao, Y. Lin, Y. Xu, W. Chen and Z. Wang, “Learning-based quality assessment for image super-resolution,” {\it IEEE Trans. Multimedia}, 2021 (Early Access).
	\bibitem{44} Y. -F. Ou, Y. Xue and Y. Wang, “Q-STAR: A Perceptual Video Quality Model Considering Impact of Spatial, Temporal, and Amplitude Resolutions,” {\it IEEE Trans. Image Process.}, vol. 23, no. 6, pp. 2473-2486, Jun. 2014.
	\bibitem{45} Z. Ma, M. Xu, Y. Ou and Y. Wang, “Modeling of Rate and Perceptual Quality of Compressed Video as Functions of Frame Rate and Quantization Stepsize and Its Applications,” {\it IEEE Trans. Circuits Syst. Video Technol.}, vol. 22, no. 5, pp. 671-682, May. 2012.
	\bibitem{47} Methodology for the Subjective Assessment of the Quality of Television Pictures, document ITU-T P.910, International Telecommunication Union, 1999.
	\bibitem{31} R. Yang, F. Mentzer, L. Van Gool and R. Timofte, “Learning for Video Compression With Hierarchical Quality and Recurrent Enhancement,” in {\it Proc. IEEE/CVF Conf. Comput. Vis. Pattern Recognit. (CVPR)}, 2020, pp. 6627-6636.
	\bibitem{32} Methodology for the Subjective Assessment of the Quality of Television
	Pictures, document ITU-R BT.500-13, International Telecommunication Union, 2012. 
	\bibitem{48} J. Yang, C. Ji, B. Jiang, W. Lu and Q. Meng, “No Reference Quality Assessment of Stereo Video Based on Saliency and Sparsity,” {\it IEEE Trans. Broadcast.}, vol. 64, no. 2, pp. 341-353, Jun. 2018.
	\bibitem{34} S. Xie, Z. Tu, “Holistically-nested edge detection,” {\it Int. Journal of Computer Vision}, vol. 125, pp. 1-3, 2017.
	\bibitem{35} J. Carreira and A. Zisserman, “Quo Vadis, Action Recognition? A New Model and the Kinetics Dataset,” in {\it Proc. IEEE Conf. Comput. Vis. Pattern Recognit. (CVPR)}, 2017, pp. 4724-4733.
	\bibitem{36} Z. Wang, E. P. Simoncelli and A. C. Bovik, “Multiscale structural similarity for image quality assessment,” in {\it Proc. 37th Asilomar Conf. on Signals, Systems \& Computers}, 2003, vol. 2, pp. 1398-1402.
	\bibitem{37} M. A. Aabed, G. Kwon and G. AlRegib, “Power of tempospatially unified spectral density for perceptual video quality assessment,” in {\it Proc. IEEE Int. Conf. Multimedia and Expo (ICME)}, 2017, pp. 1476-1481.
	\bibitem{39} A. Mittal, R. Soundararajan and A. C. Bovik, “Making a “Completely Blind” Image Quality Analyzer,” {\it IEEE Signal Process. Lett.}, vol. 20, no. 3, pp. 209-212, Mar. 2013.
	\bibitem{40} J. Korhonen, “Two-Level Approach for No-Reference Consumer Video Quality Assessment,” {\it IEEE Trans. Image Process.}, vol. 28, no. 12, pp. 5923-5938, Dec. 2019.
	\bibitem{41} D. Li, T. Jiang and M. Jiang, “Unified quality assessment of in-the-wild videos with mixed datasets training,” {\it Int. Journal of Computer Vision}, vol. 129, no. 04, pp. 1238-1257, Jan. 2021. 
	\bibitem{46} W. Shen, M. Zhou, X. Liao {\it et al.}, “An End-to-End No-Reference Video Quality Assessment Method With Hierarchical Spatiotemporal Feature Representation,” {\it IEEE Trans. Broadcast.}, Apr. 2022 (Early Access).


	
	
	
	
	
	
\end{thebibliography}
\end{document}